# Click Through Rate Prediction for Contextual Advertisment Using Linear Regression

Muhammad Junaid Effendi
Dept. of Computer & Information Systems Engineering,
N.E.D University of Engineering & Technology,
Karachi, Pakistan
junaidfnd@gmail.com

Syed Abbas Ali
Dept. of Computer & Information Systems Engineering,
N.E.D University of Engineering & Technology,
Karachi, Pakistan
saaj@neduet.edu.pk

*Abstract*—This research presents an innovative and unique way of solving the advertisement prediction problem which is considered as a learning problem over the past several years. Online advertising is a multi-billion-dollar industry and is growing every year with a rapid pace. The goal of this research is to enhance click through rate of the contextual advertisements using Linear Regression. In order to address this problem, a new technique propose in this paper to predict the CTR which will increase the overall revenue of the system by serving the advertisements more suitable to the viewers with the help of feature extraction and displaying the advertisements based on context of the publishers. The important steps include the data collection, feature extraction, CTR prediction and advertisement serving. The statistical results obtained from the dynamically used technique show an efficient outcome by fitting the data close to perfection for the LR technique using optimized feature selection.

*Keywords-Click Through Rate(CTR), Contextual Advertisements, Machine Learning, Web advertisements, Regression Problem.*

## I. INTRODUCTION AND RELATED WORK

The online advertising is a very vast industry having more than 50 billion worth. The online advertisers are now growing more and more because of the targeted advertising. This field has already produced a lot of research work in the past from ad click prediction to ad serving. With the rapid increase of ad networks the problem of click prediction has also grown. The advertisement serving is considered as one of the most successful story in the field of Machine Learning. Furthermore, the rise in advertisement serving technologies have also brought a real time bidding solution where advertisements are selected based on the features of the publishers and the viewers. The CTR prediction has been used over the past several years in every type of advertisement format, search engine advertisements, contextual advertisements, text advertisements, display banner advertisements, video advertisements etc.

Machine Learning has played an important part in the online serving of advertisements, a paper presented by [1] explains how important the machine learning is in the real world. A lot of research work has been seen in this field but none has been done in the way we have explained in this paper. The methodology and the technique used brings a variation to the ad click prediction technique. The change in methodology is the change in the steps, for example the feature extraction. This step has been done in wide variety of ways for the regression problems, linear discriminant analysis (LDA) which is used for classification problem was modified to solve regressions problems by [2]. The mapping of ordinal data into a meaningful continuous stream for linear regression to avoid fitting problems in this research has been done using the concept of k-mean clustering and association rule as presented by [14], [15], respectively. The clustering and rule mining for textual categorization has been previously researched by [16], [17] but have not used to solve the regression problems. Whereas [7] presented the conversion of ordinal data into nominal data and ignored the continuous variables for the regression. On the other side, we have seen several approaches to solve the ad click prediction. It has been done using Logistic Regression by various researchers such as [3], further naïve Bayes has played a vital part in the building of ad click prediction, this can be seen in [4], [5]. With the increase in rapid growth of online activity, the researchers have been trying to find the various techniques to get the suitable advertisements according to the viewers' interest. The estimation of response for the display advertisements is presented by [6]. Similarly, the more the viewers the ad serving becomes hard to solve, the research presented by [23] has focused on the targeting display advertisements using massive-scale machine learning. Another research in this filed was done by [8] in which the focus is on the comparison of different machine Learning techniques including the logistic regression. The recent growth in the social networking on the internet has produced some great examples of ad serving technologies, Twitter advertising model is also based on the logistic regression, [20]. While, another top network Facebook has used decision trees with logistic regression presented in [21] to serve the ads to their millions of users daily. Logistic Regression has been playing a vital role since the first research in this field. The importance of logistic regression in finding the customers behavior can be seen in the paper written by [9]. These types of models have always come up with an accuracy issues, the goal is to perform regression in a dynamic way for achieving the best possible accuracy. Estimating the errors and accuracy after model selection is one of the important parts of it according to [19]. The focus of this research is centered only to the display banner advertisements which lie within the contextual type. This research paper discusses the complete methodology of the advertisement prediction with several steps which are defined in the later sections. Section II explains the basic working of an advertisement server, how the advertisements are serving over the internet in an efficient way. Section III deals with how the data is collected from viewer's activity. Section IV describes the research methodology and technique used to predict click through rate. Experimental Results and performance evaluation is discussed in section V. Finally, conclusion is drawn in section VI.

## II. BRIEF OVERVIEW OF SYSTEM

When a viewer visits any website which is publishing the advertisements, the features are extracted from the website, the content understanding through keyword gathering, ad size, ad placement (above the fold or below the fold), viewer's location, and many other important features. They are then fed into the ad server where the data processing takes place from where the ad is chosen based on several factors, these factors are usually the order of advertisements which is shown to the viewer in the past, the behavior of the viewer extracted from the browser and most importantly the click through rate of the advertisements are going to be served. This can be understood easily with the help of fig 1. This whole process needs to be done in few milliseconds else the viewer might end up leaving the website without seeing the ad in the ad space. Further feature like ad placement plays a vital role because an advertisement is not serving the purpose if it is unreachable for the viewer [24].

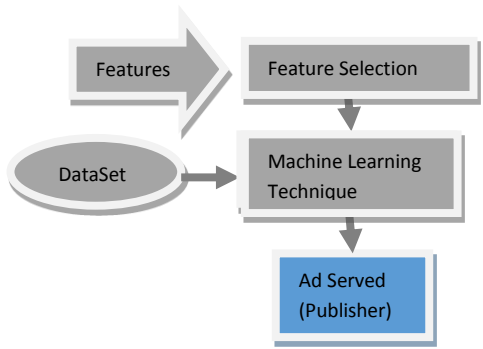

Figure 1. THE BASIC FLOW OF THE SYSTEM

The past researches have shown that this is a regression problem and perfectly fits to the logistic model [3, 8], however our goal is to use linear regression with a little variance in the feature selection method where we will use techniques like clustering to make the feature meaningful in order to avoid over fitting and under fitting problems.

## III. COLLECTION OF VIEWERS' ACTIVITY

This section presents the technique used for gathering the viewer activity which is used in later machine learning for predicting the click through rate of the advertisements. Since the collection is a necessary step of predicting the CTR for the future ads, a simple algorithm is built to serve the ads and collect the data in the raw form through our practically implemented project and the data gathered is processed later.

Following is the pseudo code for the algorithm used.

**Algorithm 1: Contextual Ad Serving based on highest bid**

Inputs: ad placement, ad size, location, keywords, category

**For** i From 1 To N do

    **If** (ad placement && ad size && location && keywords && category)

    *//setting advertisements pool*

    Adsi <- **Ad**

    **End If**

**End For**

**For All i in Ads do**

    **For All j in Ads do**

    *//preferring content based*

    Adi <- Adsi (count(keywords)==max(keywords))

        **If** Adj count(keywords) == Adi count(keywords)

    *//getting the ad with maximum bid from the pool*

    Ad <- Adi (bid==max(bid))

    **End If**

    Display <- Ad *//display the selected Ad*

    **End For**

**End For**

The purpose of this algorithm is to pick the ad based on the context of the publisher plus the maximum bidder within that contextually selected pool of advertisements. In simpler words, it prefers an ad which is more relevant to the content and then picking the highest bidder in the contextually selected pool of ads. Thus the data collected through this is used in the next sections to perform Linear Regression.

## IV. RESEARCH METHODOLOGY AND TECHNIQUES

There are several ways to perform ad click prediction, the method and techniques may vary among different types of data sets. From the fig 1 discussed in section III, we have modified the feature selection and machine learning technique blocks using different techniques to serve the same purpose. This section covers the main method and technique used to predict the click through rate of the ad. This will be explained step by step with a proper prove in a proper manner.

### A. Feature Extraction And Selection

This section deals with two types of features, one which is gathered in the previous section and used to serve advertisements using a contextual and bidding algorithm and the second one extracted on the run time from the viewer, for which the system is going to predict a CTR. Whenever a data is gathered through run time it has several features, the first task is to get the required set of features which are best for the learning algorithm. The data obtained is not in a suitable form before processing it must be converted to avoid regression problems like over fitting and under fitting [11]. Since there are three different entities Advertiser, Publisher and Viewer linked, so many features are extracted. The following table shows the features extracted from three different sources.

Table 1.  COMPARATIVE ANALYSIS OF FEATURES USING MACHINE LEARNING CLASSIFIERS

| | Features |
|---|---|
| Advertiser | Advertisement Id, Campaign Id, Category, Ad Size, Bid, landing page |
| Publisher | Ad Placement, Content (Keywords) |
| Viewer/User | Location (Area, City, Country), IP, Browser, Cookies |

These above features can be categorical, nominal and ordinal, also can be discrete or continuous in nature. The features extracted are now needed to be selected efficiently for the learning purpose. The above table has many features that are irrelevant for learning and has no meaning to the click through rate, the output. Advertisement Id, Campaign Id and several other features have no impact on the CTR, so they can be removed from the advertiser list. For the publisher part, all features are important despite that the keywords are meaningless for the regression problems but our task is to make it meaningful since its an important feature for contextual advertisements. This is discussed in the next part of the current section. The features extracted from user are important but not in this scenario because of the toughness of the regression problems, adding them would make the model over fit, thus reducing the efficiency of the system. Following table shows the feature selected for Linear Regression.

Table 2.  FEATURES SELECTED FOR REGRESSION

| | Features |
|---|---|
| Advertiser | Ad Size, Bid (The floor price) |
| Publisher | Ad Placement, Content (Keywords), |

The above features selected now need to be converted into a suitable form, a form that can be easily processed for the Regression Learning algorithm. Since Regression works only with continuous values, bid remains a perfect feature so no need to work on it. But features like ad size, keywords, ad placement have to be changed into a proper form.

Ad Size and Ad Placement are categorical features these can be mapped as:

- Ad Size to integer, '300x250' to 1, as an example.
- Ad Placement to binary, 'Above the Fold' to 1, as an example.

But keywords are values which need to be converted in a continuous stream, since there is no limit and any keyword can be generated by advertiser. The next part focuses on the keyword conversion to continuous stream using clustering to make them meaningful and impactful on the output variable CTR. At this moment the data gathered has the value for CTR, the features and the output now can be used to predict the CTR for the new viewer. For the second type of feature selection where CTR is to be predicted the same process discussed earlier takes place with same set of features as shown in table 2 but used to find the click prediction of the ad. This will be used in the later part of this section.

*B. Keywords To Integer Mapping Using Association Rule*

The next step is to make the feature keywords meaningful and impactful on the data. As discussed the keywords cannot be converted into a numerical value because of its random generation. Taking as an example, we have a keyword 'football' and we have to map it to a numerical value which is assumed to be 21. For the next keyword 'soccer' this can however be mapped to a value double to the value of the 'football' keyword 42 or any other value which is not repeated in the system, but the evaluation here will not make any useful impact and will be of meaningless value, this would lead to over fitting problem. In order to solve this problem Association Rule can be used to make it meaningful and to avoid fitting problems [12]. This feature can also be ignored but to add a different dynamic to the problem we have considered it as an important part of learning. Also, because the focus is on the contextual advertisements so keywords here define the context of the page on which the ad needs to be predicted. The Rule Mining and clustering technique is used to understand the given keywords would benefit us by avoiding the over-fitting problem without the use of regularization. The categories of the advertisements are labeled by default and the keywords within it will be mapped into a meaning numerical with respect to its distance.

Table 3.  LABELLED CATEGORIES

| Categories |
|---|
| Sports |
| Health |
| Fashion |

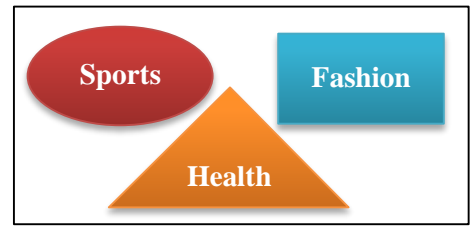

Figure 2.  THE DEFAULT CLUSTERS

The focus is only on the sports category and all analysis are based on this category. Following are the sample set of keywords from the sports category.

Table 4.  UNLABELED KEYWORDS FOR CATEGORY SPORTS

| Football | La Liga |
|---|---|
| Soccer | Real Madrid |
| Premier League | Afridi |
| Tennis | Ronaldo |
| Cricket | EPL |
| Nadal | Spain |
| Brazil | Federer |
| Pakistan | England |

At the moment these keywords look like in the following manner.

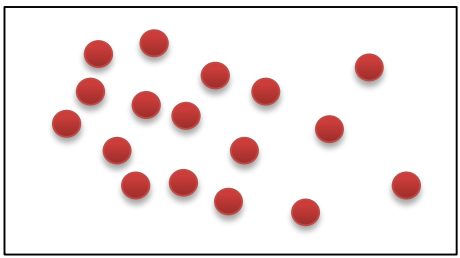

Figure 3. THE DEFAULT CLUSTERS

The three different colors (purple, yellow, blue) shown in fig 4 are the three centroids selected; they are the keywords that represent the clusters, for example the cricket. And the centroids are found using the association rule on the data set collected in the starting section of this research paper.

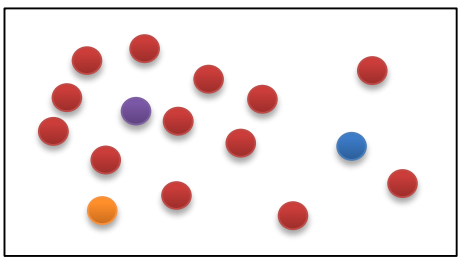

Figure 4. SHOWING THE MARKING OF THREE CENTROIDS

The centroids are picked as the most commonly used keywords. For example, football, cricket and tennis are the most common terms used with other keywords.

The association rule mining is a technique to find the relationship among certain data set [16]. With the help of this technique the categorization can be done efficiently until and unless the data set has those keywords.

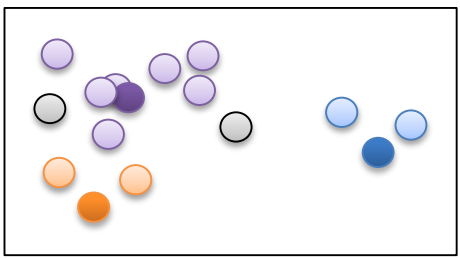

Figure 5. MAKING CLUSTERS (PURPLE FOR FOOTBALL, YELLOW FOR CRICKET, BLUE FOR TENNIS)

The dark colored are the centroids while the light colors are the related keywords. The overlapping in the purple part of two circles is because of the closeness between the soccer and football keywords. The two gray are the neutral keywords that can be categorized in more than one cluster. England and Spain keywords here refer to those two. Spain can be in football or tennis category and England can be in football or in cricket. The following Fig.6, focus only on the purple clusters, football. Also considering the two neutral circles in its category.

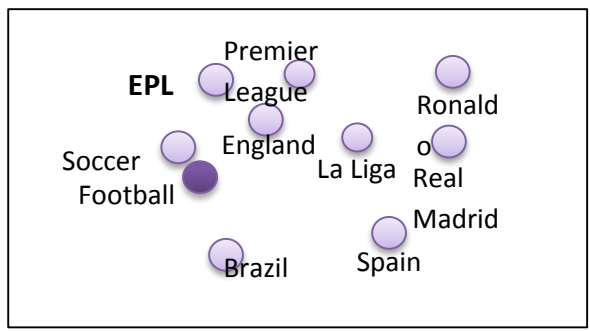

Figure 6. A CLOSER LOOK OF PURPLE CLUSTERS

Here the distance is showing the relation among the keywords. In the real world, Football or soccer is linked with England then it has its Premier League also known as EPL, the La Liga is the Spanish League having a club Real Madrid with its player Ronaldo. Brazil is a direct link to football. The closeness between the keywords can be defined in terms of the distance between the two keywords. The more related the keywords the less the distance between them. This can be viewed in the fig 6. Now the distance between them can be used to map them into a meaningful numerical value. As a sample the following table shows result.

Table 5. MAPPING OF KEYWORDS TO NUMERICAL VALUE

| Football | |
|---|---|
| Football | 50 |
| Soccer | 49.9 |
| England | 51 |
| Premier League | 52 |
| EPL | 52.1 |
| Spain | 49 |
| La Liga | 48 |
| Real Madrid | 47 |
| Ronaldo | 47.5 |
| Brazil | 49.5 |

The above table shows how much each keyword is related to other in terms of the distance value. The more the keywords the closer the value would be. The values may differ each time since this technique performs every time when the learning needs to be done.

*C. CTR Prediction Using Machine Learning*

Now all the features have been converted into a suitable form and learning can be done easily. The click through rate is predicted when a viewer visits a publisher's website and within few milliseconds the ad is being served based on the maximum CTR. The features as discussed in the previous section are now sent to calculate the CTR of the ads. The ads are picked according to the location of the viewer and category of the website. The problem has been carried out in two ways, the linear regression equation and the normal equation presented by [10]. The predictions are performed using both the ways and are compared in the next part of this section.

A) Data Set with Trend

The data gathered in the earlier section can be shown in the form of graph given below:

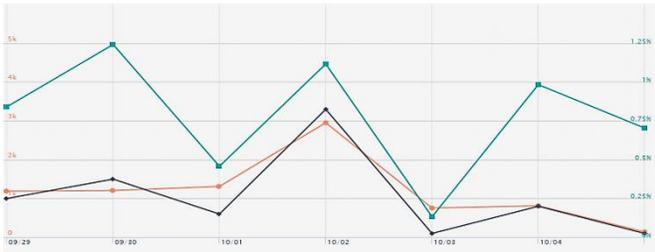

Figure 7. PLOTTING OF AD IMPRESSIONS (ORANGE), CLICKS (BLUE) AND CTR (LIGHT BLUE)

The fig 7 data have been collected from the last week of September; 2016. The following detailed form is extracted with the help of figure 7, showing only 12 records as a sample.

Table 6. TABLE 6: SUITABLE FORM OF FEATURES AS DISCUSSED IN THE PREVIOUS PART

| Ad Placement | Ad Size | Bid | Keywords (Table 5) | CTR |
|---|---|---|---|---|
| 1 | 1 | 20 | 50 | 0.08 |
| 1 | 1 | 15 | 47 | 0.04 |
| 1 | 2 | 10 | 52.1 | 0.02 |
| 1 | 2 | 40 | 52 | 0.06 |
| 0 | 3 | 20 | 49.9 | 0.01 |
| 0 | 1 | 15 | 49.5 | 0.006 |
| 0 | 1 | 10 | 52.1 | 0.005 |
| 1 | 1 | 42 | 52 | 0.1 |
| 0 | 1 | 25 | 47.5 | 0.015 |
| 1 | 2 | 20 | 47 | 0.02 |
| 1 | 2 | 10 | 48 | 0.001 |
| 0 | 3 | 5 | 50 | 0.0001 |

B) Linear Regression

As discussed Linear Regression works only on continuous values of both independent and dependent variables.

**The formula of single input LR is given by:**

$$h_\theta(x) = \theta_0 + \theta_1 x_1 \quad (1)$$

The formula of multiple inputs LR is given by:

$$h_\theta(x) = \theta_0 + \theta_1 x_1 + \theta_2 x_2 + \cdots + \theta_n x_n \quad (2)$$

$$h_\theta(X) = \theta^T X$$

Before proceeding the data needs to be normalized because of the expected gap among the values of a single feature especially the keywords, 'football' and 'Ronaldo' might have a gap which would reduce the accuracy, so feature scaling, in other words normalization could solve this problem. Mean and Standard Deviation are the vital two terms that are part of normalization.

Mean:

$$\mu = \frac{\sum x}{n-1} \quad (3)$$

Standard Deviation:

$$\sigma = \sqrt{\frac{\sum(x-\bar{x})^2}{n-1}} \quad (4)$$

Normalization formula:

$$x_{new} = \frac{x - \mu}{\sigma} \quad (5)$$

Our goal is to calculate the appropriate parameters to predict the out, the formula to calculate the parameters is given by:

$$J(\theta_0, \theta_1) = \frac{1}{2m} \sum_{i=1}^{m}(h_\theta(x^{(i)}) - y^{(i)})^2 \quad (6)$$

The main task is to reduce the cost by minimizing the parameters:

$$\underset{\theta_0, \theta_1}{\text{minimize}} J(\theta_0, \theta_1)$$

To make it efficient the cost must be reduced using the gradient descent:

Repeat until convergence {

$$\theta_j := \theta_j - \alpha \frac{\partial y}{\partial \theta_j} J(\theta_0, \theta_1) \quad (7)$$

(for j = 1 and j = 0)

This must run until the parameters start converging. The Simple Linear Regression also known as the single input/feature/independent variable linear regression is used to have the insights of the one to one relationship of each feature with the output. The $\theta_0$ in both types of linear regression is the interception value. This value is set to 1 which is passed as the interception value, often called as constant [25]. It is used to found the mean of y in case of x=0, however in our problem the intercept value can be removed because the predictors will never be a zero value. It is the response or the output value when all parameters are set to zero [11], [12].

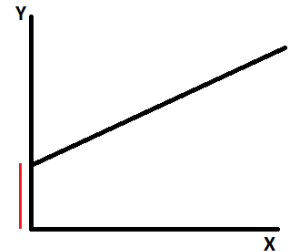

Figure 8. THE RED LINE SHOWING THE INTERCEPT VALUE

Alpha ($\alpha$) in the gradient descent function is the learning rate which helps alongside the number of iteration to find the best possible parameters. The greater the value of $\alpha$ the less iterations would be required and vice versa [13]. Here $\alpha$ is set to 0.01 and number of iterations to 400. The above two values are checked and tested. It gives the optimal value of error.

## V. EXPERIMENTAL RESULTS AND PERFORMANCE EVALUATION

This section illustrates the demonstrative experiments to enhance click through rate of the contextual advertisements using Linear Regression based on context of the publishers and evaluate the performance of proposed technique in term of accuracy and standard error.

1. Simple Linear Regression

The advertisement size and placement are categorical in natural so they can be ignored for simple linear regression. But all continuous based features like bid and keywords are used to show the direct relation with the outcome CTR using a graphical view. The simple linear regression has been defined earlier in this section is used to find the relationship for the following two relations. The data used has been converted into a normalized form.

- Relationship of Bid with CTR

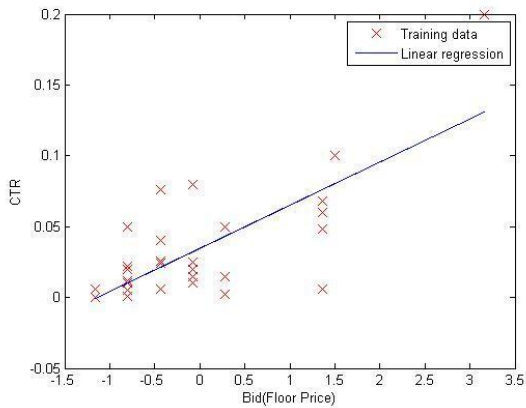

Figure 9. BID AND CTR RELATION THROUGH LINEAR REGRESSION LINE

Fig 9 shows a linear relation between bid and click through rate. The increase in bid means an increase in CTR and vice versa. Parameters found: 0.034421, 0.030570. For bid = 22, the predicted CTR is 0.036440

- Relationship of Keywords with CTR

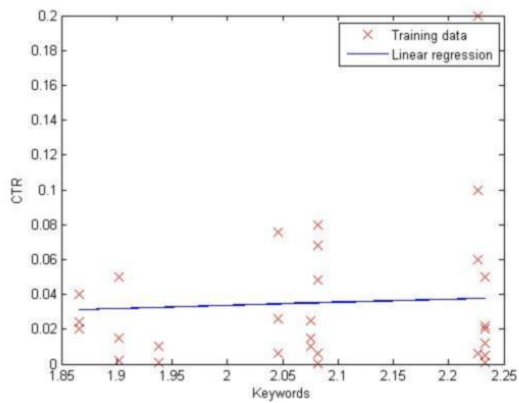

Figure 10. KEYWORDS AND CTR RELATION THROUGH LINEAR REGRESSION LINE

Fig 10 shows that there is no such relation between the keywords and CTR. The output can decrease with the increase in input and vice versa. It is an overfitting of data since many points are far from the regression line.

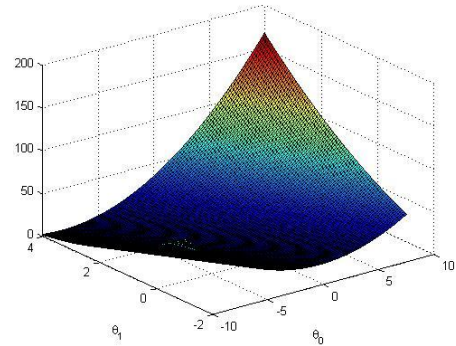

Figure 11. PARAMETERS RELATION WITH COST IN 3D

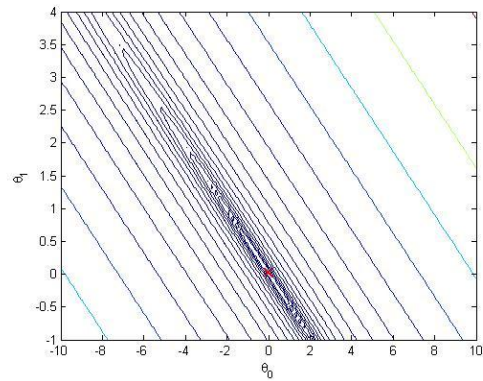

Figure 12. PARAMETERS RELATION WITH COST IN 2D

Reducing the three-dimensional space into two-dimensional, shown in fig 12.

Parameters found: -0.001484, 0.017645

For Keyword = 51 (England), the predicted CTR is 0.009720. This could be a negative value upon changing the keyword. From fig 10 it is proved that the keyword relation with CTR is not typical as of previous relation but somehow using this feature alongside other features gives an efficient result, discussed in the next part.

2. Multivariate Linear Regression

Now taking all features into account and using all in a multivariate linear regression to predict the outcome.

The data which has been normalized is used to find the parameters using the gradient descent. Parameters calculated are shown in the table 7. Fig 13 explains how cost function is converged with respect to the number of iteration which is set to 400.

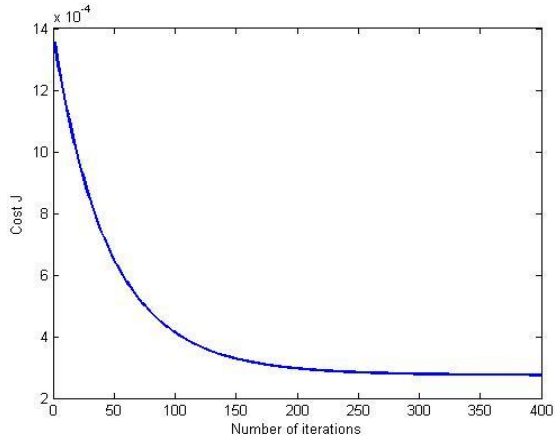

Figure 13. THE BLUE LINE SHOWING THE DECREASE IN COST

From Matlab, the following results are obtained;

Table 7. LR PREDICTED CTR

| Inputs | | Parameters | Output |
|---|---|---|---|
| Intercept | 1 | 0.033804 | |
| Placement | 1 | 0.010290 | The predicted |
| Size | 1 | 0.000785 | Click through |
| Bid | 22 | 0.027315 | Rate is 0.048436 |
| Keyword | 51 | 0.005490 | |

The LR using feature scaling and gradient descent gives a result 0.042821. Normal Equation is an easy way of solving Linear Regression problems. It does not require the steps as performed above. But the limitation is that it cannot be used for a large feature set. Normal equation is given by:

$$\theta = (X^T X)^{-1} X^T \vec{y} \quad (8)$$

From Matlab, the following results are obtained;

Table 8. NORMAL EQUATION PREDICTED CTR

| Inputs | | Parameters | Output |
|---|---|---|---|
| Intercept | 1 | -0.157028 | |
| Placement | 1 | 0.020795 | The predicted |
| Size | 1 | 0.002185 | Click through |
| Bid | 22 | 0.002040 | Rate is |
| Keyword | 51 | 0.002696 | 0.048338 |

The normal equation gives a result 0.042946, almost equal to that we got from the previous method but with no work like cost reduction and feature scaling. This result will vary by changing the no. of records in the data set. The results obtained using the keyword as a feature gives almost the same result from the two ways, linear regression and normal equation. The accuracy and the standard error of the system is calculated using the cross validation set, given in table 9:

Table 9. CROSS VALIDATION SET

| Ad Placement | Ad Size | Bid | Keywords (Table 5) | CTR |
|---|---|---|---|---|
| 1 | 1 | 12 | 52 | 0.03 |
| 1 | 2 | 32 | 51 | 0.05 |
| 1 | 1 | 14 | 49 | 0.042 |
| 1 | 1 | 21 | 50 | 0.04 |
| 0 | 2 | 25 | 47.1 | 0.0001 |
| 0 | 3 | 6 | 48 | 0.00005 |

Standard Error:

$$SE = \sqrt{\frac{\Sigma(y-y')^2}{N}} \quad (9)$$

y is the observed CTR, $y'$ is the predicted CTR and N is the total no. of records which is 6.

Table 10. CALCULATION OF STANDARD ERROR

| y | y' | f= y-y' | Squared f |
|---|---|---|---|
| 0.03 | 0.031575 | -0.00158 | 2.48E-06 |
| 0.05 | 0.069137 | -0.01914 | 0.000366 |
| 0.042 | 0.052655 | -0.01066 | 0.000114 |
| 0.04 | 0.033828 | 0.006172 | 3.81E-05 |
| 0.0001 | -0.00041 | 0.000507 | 2.57E-07 |
| 0.00005 | -0.00968 | 0.009731 | 9.47E-05 |
| **0.16215** | **0.029517833** | **-0.01496** | **0.000615** |
| **ym:0.027025** | | | |

The second last row is the sum and the last row is the mean of the respective column. Standard Error found to be equal to 0.010126512. These types of errors have a direct effect on the accuracy of the training and test data and this is usually found in linear regression [18]. R squared is a useful statistical measure which helps in finding the accuracy of the model [19].

R squared:

$$R^2 = 1 - \frac{\sum_{i=1}^{n}(y_i - y')^2}{\sum_{i=1}^{n}(y_i - ym)^2} = 1 - \frac{SSE}{SSTO} \quad (10)$$

SSE is the sum of squares error, SSTO is the total sum of squares.

Table 11. CALCULATION OF R SQUARED

| Squared f (table 9) | f1=y-ym | squared f1 |
|---|---|---|
| 2.48E-06 | -0.01221 | 0.000149 |
| 0.000366 | 0.007792 | 6.07E-05 |
| 0.000114 | -0.00021 | 4.33E-08 |
| 3.81E-05 | -0.00221 | 4.88E-06 |
| 2.57E-07 | -0.04211 | 0.001773 |
| 9.47E-05 | -0.04216 | 0.001777 |
| **0.000615** | | **0.003765** |

The last row is the mean value of the respective column. R squared found to be equal to 0.836581861, meaning that 83.65% of data has fitted the model correctly. The accuracy can be increased by removing the keyword from the feature set but the concern here is to use the dynamically converted keyword into a meaningful value as a feature to obtain the best possible results.

VI. CONCLUSION

This research presented a new technique of predicting the click through rate for online advertisements using Linear Regression along with some dynamically added feature known as the keyword. The proposed technique helps to calculate the CTR from a different angle despite causing a minor decrease in efficiency. The accuracy found is 83% and removing that feature could take this accuracy to 95% which is a significant increase and fits the model perfectly but the CTR is also dependent on the keywords since the research is based on the contextual advertisements. For future research work, these results can help to combine and reveal further more techniques to enhance the performance of the advertisement serving over the internet.